\documentclass[twocolumn,preprintnumbers,amsmath,amssymb,showkeys,mathtools,letterpaper]{revtex4-1}
\pdfoutput=1
\usepackage{graphicx}
\usepackage{upgreek}
\usepackage{amsmath}
\usepackage{amssymb}
\usepackage{longtable}
\usepackage{ctable}
\usepackage{graphicx}
\usepackage{amssymb} 
\usepackage{color}

\begin{document}

\title{Long-Range Interactions Dominate the Inverse-Temperature Dependence of Polypeptide Hydration Free Energies}

\author{Dheeraj S. Tomar}
\affiliation{Chemical and Biomolecular Engineering, Johns Hopkins University, Baltimore, MD 21208}
\affiliation{Akrevia Therapeutics, Cambridge, MA}
\author{Michael E. Paulaitis} 
\affiliation{Center for Nanomedicine, Johns Hopkins School of Medicine, Baltimore, MD 21205}
\author{Lawrence R. Pratt}
\affiliation{Department of Chemical and Biomolecular Engineering, Tulane University, New Orleans, LA 70118}
\author{D. Asthagiri}\email{To whom correspondence should be addressed. E-mail: dna6@rice.edu}
\affiliation{Sealy Center for Structural Biology and Molecular Biophysics, University of Texas Medical Branch, Galveston, TX 77555}
\affiliation{Department of Chemical and Biomolecular Engineering, Rice University, Houston, TX 77005}

\date{\today}


\begin{abstract}
Direct, all-atom calculations of the free energy of hydration
of aqueous deca-alanine structures --- holistically including 
backbone and side-chain interactions together --- show that attractive
interactions and the thermal expansion of the solvent explain 
the inverse temperature signatures that  have been  interpreted traditionally
in favor of hydrophobic mechanisms for stabilizing the structure and function 
of soluble proteins. 
\end{abstract}

\maketitle

\section{Significance Statement}
Solution environments of soluble proteins are intrinsic to the thermodynamic stability of those aqueous macromolecular species. Only 
recently has molecular theory and simulation progressed to the stage that hydration free energies of soluble proteins can be evaluated 
holistically, including peptide backbone moieties at the same level as side-chain groups.  The new results provide surprising insight into inverse temperature 
dependences that have been implicated in cold denaturation of these structures.  Thus, these results should change our qualitative appreciation of the solution influence on the stability and function of biomolecular structures.

\section{Introduction}
Solution environments of soluble proteins are intrinsic to the thermodynamic stability of those macromolecular structures.  Typically, a protein is divided into hydrophilic and hydrophobic moieties, then characteristic hydration free energy contributions are assigned, and assembled additively, to rationalize their solution structure, stability, and function.  The hydrophilic/hydrophobic assignment is often somewhat arbitrary.  But distinctive temperature dependences --- so-called inverse temperature  dependences \cite{doi:10.1021/acs.jpcb.8b01711} --- that are characteristic of hydrophobic free energies support the hydrophilic/hydrophobic dichotomy.   Strengthening of hydrophobic stabilization with increasing temperature in a physiological range is a simple example of inverse temperature behavior, and viewed in the decreasing temperature direction rationalizes cold denaturation \cite{franks1988thermodynamics}.

Nevertheless, Klotz pointed-out some time ago \cite{klotz1999parallel} that inverse temperature behavior can be observed in aqueous chemical equilibria 
such as simple carboxylic acid dissociation involving highly hydro\emph{philic} species. Knowledge of the hydration thermodynamics of small molecule analogues of groups
comprising the protein indeed do frame views of the dominant forces in protein folding \cite{Kauzmann:59,tanford:62}.  Though molecular simulations have played a decisive role in
filling-out this knowledge, calculating the hydration thermodynamics of a protein holistically at the level that is possible for, say, CH$_4$ has remained unaddressed.
With developments in the molecular quasi-chemical theory \cite{lrp:ES99,lrp:apc02,lrp:book,lrp:cpms,safirphd} of solutions and the associated simulation implementation
\cite{weber:jcp10b,weber:jcp11}, this situation has changed \cite{Weber:jctc12}. Refinements in the simulation methodology \cite{tomar:bj2013,tomar:jpcb14} have made it possible to
interrogate the thermodynamics of macromolecules at a level that has only been undertaken for small solutes \cite{tomar:jpcb16,asthagiri:gly15,tomar:gdmjcp18} Here we bring those 
tools to bear on the temperature dependence of the hydration thermodynamics of a polypeptide, thus characterizing the thermodynamic forces driving protein folding. 

We study the hydration thermodynamics of the deca-alanine polypeptide in helical and an extended coil conformation. The coil conformation, labeled $C_0$, has the least negative hydration free energy of the coil states studied earlier \cite{tomar:jpcb16}, bounding the free energy of the unfolded ensemble from above \cite{asthagiri:gly15}.  We also study the
thermodynamics of a pair of helices in contact. The helix pair serves as a model of protein tertiary structure. The long axis of the helices are parallel, and the helix macro-dipoles are either anti-parallel, as might occur in a helix-turn-helix motif, or parallel, as might occur in a helix bundle. (In nature helices are not perfectly aligned, but this issue is secondary to the questions studied here.) We also contrast the studies on the deca-alanine with the hydration of CH$_4$, the small-molecule analogue of the alanine side-chain. 

Our results for hydration of the poly-alanine peptides show that the sign of the partial molar excess entropy, $s^{\mathrm{(ex)}}$, and the partial molar heat capacity, $c_p^{\mathrm{(ex)}}$, are just as found for CH$_4$.  In the disassembly of the helix-pair, for example, although we recover the well-known $\Delta c_p^{\mathrm{(ex)}} > 0$ observed in
protein unfolding, this signature reflects the weakening of the effective protein-solvent attractive interactions. In a curious twist, the thermal expansion of the water matrix, implicated in 
hydrophobic interactions \cite{pratt2007special,pohorille2012water,pratt2016statistical}, also plays an important role in the temperature signatures  in the protein models.

\section{Theory} 

The calculation of $\mu^{\mathrm{(ex)}}$ and its entropic $Ts^{\mathrm{(ex)}}$ and enthalpic $h^{\mathrm{(ex)}}$ contributions follows earlier work
\cite{tomar:bj2013,tomar:jpcb14,tomar:jpcb16,asthagiri:gly15,tomar:gdmjcp18}.  Briefly   \cite{widom:jpc82,lrp:cpms,lrp:book}, the excess chemical potential is given by
$\beta \mu^{\mathrm{(ex)}} = \ln \langle e^{\beta \varepsilon} \rangle$, averaging $\varepsilon$  over the binding energy distribution $P(\varepsilon)$. As usual, $\beta =
1/k_{\rm B}T$, with $T$ the temperature and $k_{\rm B}$ the Boltzmann constant. We regularize \cite{asthagiri:jcp2008,safirphd} the
calculation of $\mu^{\mathrm{(ex)}}$ by introducing an auxiliary field $\phi(r;\lambda)$ that moves the solvent away from the solute, thereby tempering the solute-solvent binding energy.   The conditional distribution $P(\varepsilon|\phi)$ is better characterized than $P(\varepsilon)$, and in calculations we adjust $\lambda$, the range of
the field, to control approximation of $P(\varepsilon|\phi)$ as a Gaussian distribution.

\begin{figure}[h!]
\includegraphics[width=3.25in]{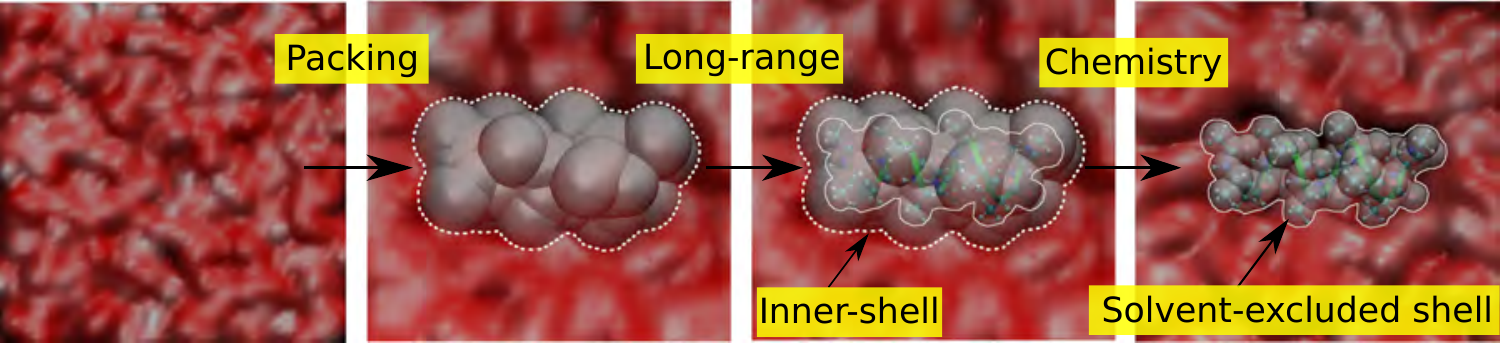}
\caption{Quasi-chemical organization of the excess chemical potential.
The inner-shell of width $\lambda_\mathrm{G} = 5$~{\AA} is
the smallest region enclosing the solute for which the
solute-solvent binding energy distribution $P(\varepsilon|\phi)$ is
accurately Gaussian. It approximately corresponds to the 
traditional first solvation shell of
the solute. The shell of width $\lambda_\mathrm{SE} \leq
3.0$~{\AA} is the envelope for which the chemistry contribution is zero, 
and thus encloses the volume excluded to the solvent. }
\label{fg:fig1}
\end{figure}

With the introduction of the field \cite{Weber:jctc12,tomar:bj2013,tomar:jpcb14,tomar:jpcb16}
\begin{eqnarray}
\beta \mu^{\mathrm{(ex)}} = 
\underbrace{- \ln p_0[\phi]}_{\rm packing} + 
\underbrace{\beta\mu^{\mathrm{(ex)}} [P(\varepsilon|\phi)]}_{\rm long-range} + 
\underbrace{\ln x_0[\phi]}_{\rm chemistry}~,
\label{eq:qc}
\end{eqnarray}
the quasi-chemical \cite{asthagiri:cpl10}  organization of the potential distribution theorem \cite{lrp:apc02,lrp:book,lrp:cpms}.  The individual
contributions are functionals of the auxiliary field, as indicated. The packing and chemistry contributions are the proximal solvent contributions that
are added back to the long-range, regularized problem to complete the calculation. Fig.~\ref{fg:fig1} provides a schematic description of
Eq.~\ref{eq:qc}.

The packing contribution measures the free energy to create a cavity to
accommodate the solute and describes primitive hydrophobic effects
\cite{Pratt:1992p3019,Pratt:2002p3001}, \emph{i.e.,}\ hydration of an ideal
hydrophobe.  The chemistry contribution captures the role of solute
attractive interactions with solvent in the hydration layer 
within a distance $\lambda$ from the center of the nearest heavy atom.  The
long-range contribution is the free energy of interaction between the
solute and the solvent when solvent is excluded from the hydration
layer.  The chemistry plus long-range contribution describes the total
hydrophilic contribution to hydration. 

The packing and chemistry contributions in these calculations are based on
a soft-cavity \cite{chempath2009quasichemical,weber:jcp11}. We find that 
$\lambda \approx 5$~{\AA}  ensures that the conditional
binding energy distribution is Gaussian to a good approximation. We
denote this range as $\lambda_\mathrm{G}$.  The largest value of $\lambda$,
labeled $\lambda_{\rm SE}$, for which the chemistry contribution is
negligible has a special meaning. It bounds the domain excluded
to the solvent.  We find
$\lambda_{\rm SE} \approx 3$~{\AA}, and emphasize that for the given forcefield and 
solute geometry, this surface is substantially unambiguous.  With this 
choice, Eq.~\ref{eq:qc} can be rearranged as, 
\begin{multline}
\beta \mu^{\mathrm{(ex)}}  =   
\underbrace{-\ln p_0(\lambda_{\rm SE})}_{\rm solvent\, exclusion} \\
+  
\underbrace{\beta\mu^{\mathrm{(ex)}}[P(\varepsilon|\lambda_\mathrm{G})]}_{\rm long-range} \\
+ 
\underbrace{ \ln \left[x_0(\lambda_\mathrm{G})\left(\frac{p_0(\lambda_{\rm SE})}{p_0(\lambda_\mathrm{G})}\right)\right]}_{\rm revised\, chemistry}~.
\label{eq:qc1}
\end{multline}
In Eq.~\ref{eq:qc1} the various contributions are
identified by the range parameter. Thus, for example,
$x_0(\lambda_\mathrm{G})\equiv x_0[\phi(\lambda_\mathrm{G})]$. 
The revised chemistry term has the following physical
meaning. It is the work done to move the solvent interface a distance
$\lambda_\mathrm{G}$ away from the solute relative to the case when the
only role played by the solute is to exclude solvent up to $\lambda_{\rm
SE}$.  This term highlights the role of \emph{short-range} solute-solvent
attractive interactions on hydration. Interestingly, the range between
$\lambda_\mathrm{SE} = 3$~{\AA} and $\lambda_\mathrm{G} = 5$~{\AA}
corresponds to the first hydration shell for a methane
carbon \cite{asthagiri:jcp2008} and is an approximate descriptor of the
first hydration shell of groups containing nitrogen and oxygen heavy
atoms. 

The excess entropy of hydration is  given by \cite{tomar:jpcb14}
\begin{eqnarray}
T  s^{\mathrm{(ex)}} &  = & 
E^{\mathrm{(ex)}} - kT^2 \alpha_p + p\left(\left\langle V^{\mathrm{(ex)}} \right\rangle + kT \kappa_T\right) - \mu^{\mathrm{(ex)}} \nonumber \\ 
	              & \approx & E_{\mathrm{sw}} + E_{\mathrm{reorg}}  - \mu^{\mathrm{(ex)}}
\label{eq:entropy}	              
\end{eqnarray}
where $\kappa_T$ is the isothermal compressibility of the solvent,
$\alpha_p$ is the thermal expansivity of the solvent,  and $\langle
V^{\mathrm{(ex)}} \rangle$ is the excess volume of hydration, and in
writing the second line of the equation, we have ignored the small
contribution from all these terms.  The average excess energy of
hydration, $E^{\mathrm{(ex)}}$,  is the sum the average solute-water
interaction energy $E_{\mathrm{sw}}$ and $E_{\mathrm{reorg}}$, the reorganization energy. 
Ignoring pressure-volume effects, the excess enthalpy of hydration
$h^{\mathrm{(ex)}} =  E^{\mathrm{(ex)}}$.

\section*{Results}

\subsection*{Hydration of Methane}
\begin{figure*}
\centering
\includegraphics{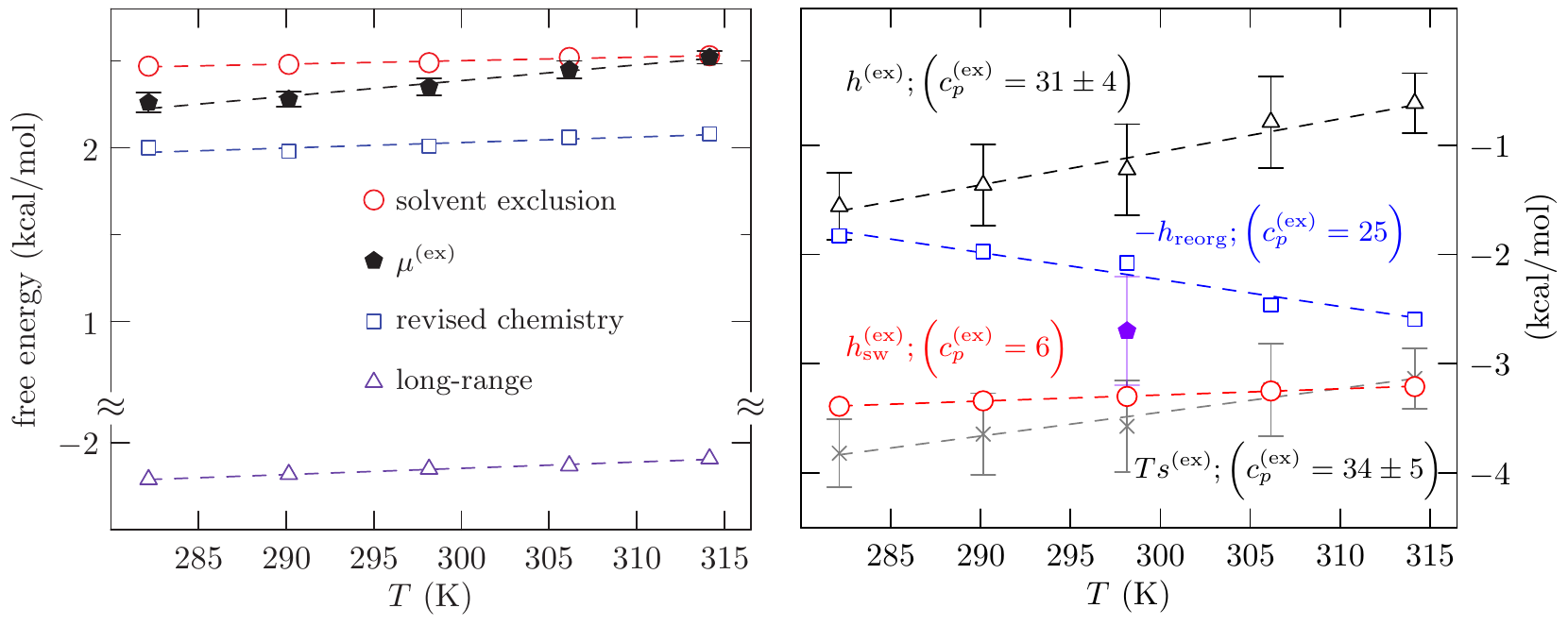}
\caption{\underline{Left panel}: The hydration free energy of methane,
following \eqref{eq:qc1}. With the
expected opposition of long-range/attractive interactions to the
other contributions, the unfavorable net hydration free energy (filled-pentagons)
increases with increasing $T$, the classic hydrophobic inverse temperature behavior. 
Note that the difference  between the unfavorable solvent exclusion contribution and the net hydration free energy decreases
with increasing $T$, showing that long-range/attractive interactions
dominate the inverse temperature behavior in this hydrophobic hydration 
phenomenon.
\underline{Right panel}:  The entropic [crosses: $Ts^{\mathrm{(ex)}}$,
\eqref{eq:entropy}]  and enthalpic [triangles: $h^{\mathrm{(ex)}}$]
contributions. The filled pentagon is
$Ts^{\mathrm{(ex)}}$ from the temperature derivative of
$\mu^{\mathrm{(ex)}}$.  Contributions
from solute-solvent interaction $h^{\mathrm{(ex)}}_{\rm sw}$ (red
circles) and solvent reorganization $h_{\rm reorg}$ combine to 
give $h^{\mathrm{(ex)}}$.  $-h_{\rm reorg}$
is plotted to show all the data on the same plot. The several
result-sets are identified by the heat capacity (in cal/mol-K) from the
corresponding contribution. Error bars are $\pm\sigma$ standard error of
the mean.}\label{fg:methane}
\end{figure*}

For CH$_4$ (Fig.~\ref{fg:methane}), $\mu^{\mathrm{(ex)}}$ and the
individual contributions [\eqref{eq:qc1}] increase with increasing
temperature. The solvent exclusion contribution makes the largest numerical
contribution to the net free energy followed by the revised chemistry
contribution. These contributions are balanced by the long-range
attractive contribution, which is favorable and thus an attractive
contribution. 

Allowing water to flood the
previously empty quasi-chemical inner-shell (Fig.~\ref{fg:fig1}),
should decrease the free energy  due to favorable
solute-solvent interactions. The (positive) sign of the revised
chemistry contribution for methane suggests that in this case, however,
the solvent is being pushed into unfavorable contact with the solute. In
another words, the solvent matrix squeezes the hydrophobe, as suggested earlier \cite{richards1991protein,Pratt:2002p775,asthagiri2007non}. 

The excess enthalpy of hydration $h^{\mathrm{(ex)}}$ has two
contributions: (a) $h^{\mathrm{(ex)}}_{\rm sw}$, arising from solute-water
interactions and (b) $h_{\rm reorg}$, arising from changes in the potential
energy of the solvent matrix upon insertion of the solute in the
solvent. The reorganization contribution is obtained using a
hydration-shell-wise summation process
\cite{Matubayasi:excess94,ashbaugh:jpc96,themis:excess00,asthagiri:jcp2008,tomar:jpcb16}. 
For methane, this contribution converges within the first shell
(Supporting Information, \textbf{SI}).  

These calculations are internally consistent. The heat
capacity at a given temperature can be obtained as either 
$c_p^{\mathrm{(ex)}}  = (\partial h^{\mathrm{(ex)}} / \partial
T)_{N,p}$ 
or $c_p^{\mathrm{(ex)}} = T (\partial s^{\mathrm{(ex)}} /
\partial T)_{N,p}$. The first is
$c_p^{\mathrm{(ex)}}[h^{\mathrm{(ex)}}]$ and the second is
$c_p^{\mathrm{(ex)}}[s^{\mathrm{(ex)}}]$ (Fig.~\ref{fg:methane}). These
analyses assume that these quantities are constant over the
temperature range considered here. Within statistical
uncertainties these values are the same from both paths.
$c_p^{\mathrm{(ex)}} \approx 31$~cal/mol-K, $h^{\mathrm{(ex)}} =
-1.2$~kcal/mol, and $s^{\mathrm{(ex)}} = -12.1$~cal/mol-K are in fair
agreement with the experimental values of about
$49$~cal/mol-K, $-2.70$~kcal/mol, and $-16$~cal/mol-K \cite{cabani:ch4}, respectively.
Deficiencies of both the solute model and
the water model impact the numerical agreement, of course; being 
consistent in these choices should be 
lead to reliable physical conclusions. 
$s^{\mathrm{(ex)}}[T = 298.15~{\rm K}]$ obtained from a
temperature derivative of $\mu^{\mathrm{(ex)}}$ agrees with the value from
\eqref{eq:entropy}, emphasizing the internal
consistency of these calculations.

\subsection*{Hydration of helix and coil conformers}
\begin{figure*}
\includegraphics[width=0.975\textwidth]{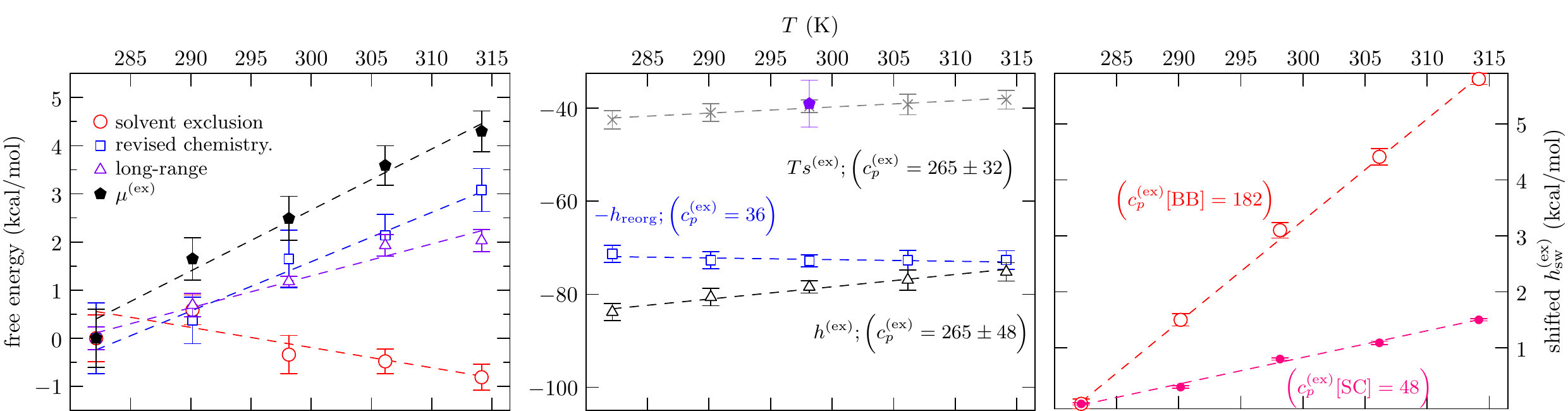} \\
\includegraphics[width=0.975\textwidth]{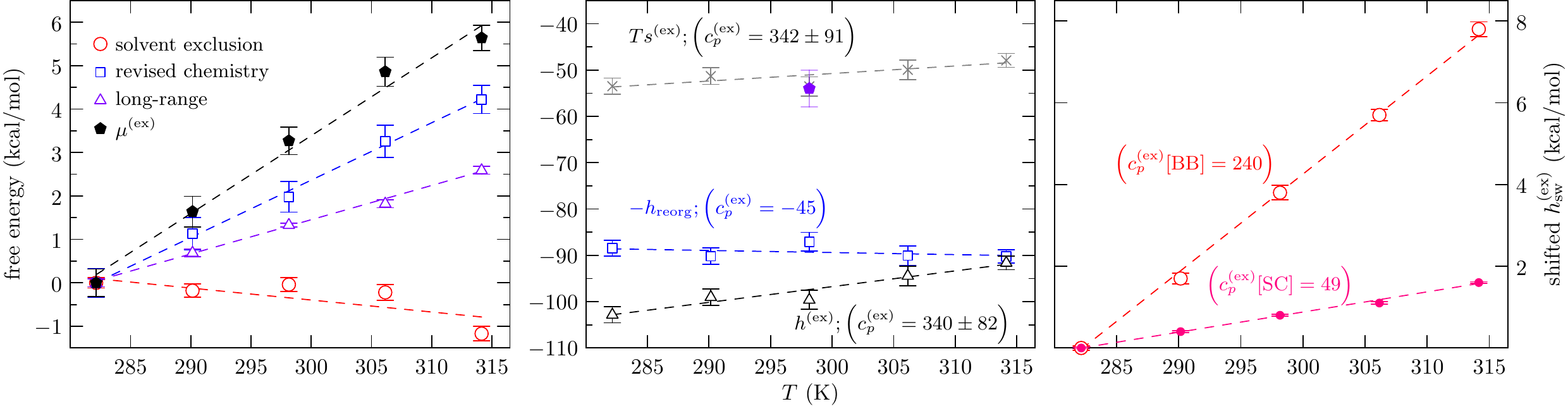} \\
\caption{As in Fig.~\ref{fg:methane} but \underline{\emph{relative}} to the 
$T = 282.15$~K value.
$h^{\mathrm{(ex)}}_{\rm sw} = h^{\mathrm{(ex)}}_{\rm BB} +
h^{\mathrm{(ex)}}_{\rm SC}$, again relative to the value at 282.15~K, is presented
separately in the rightmost panel.  $h^{\mathrm{(ex)}}_{\rm BB}$ is the
contribution from backbone-solvent interaction and
$h^{\mathrm{(ex)}}_{\rm SC}$ is the contribution from side-chain solvent
interaction. \underline{Top row}: Helix. \underline{Bottom row}: Coil.
  }\label{fg:helixcoil}
\end{figure*}

Moving to consider  hydration of the
helix and coil conformers (Fig.~\ref{fg:helixcoil}), note  the
internal consistency of those calculations. The comparatively large
uncertainty in $c_p^{\mathrm{(ex)}}$ arises from $h_{\rm
reorg}$.  The shell-wise calculation of the reorganization contribution
considerably reduces the uncertainty relative to naive direct
differencing of potential energies of the entire solvent bath with and
without the solute.   Still, it is difficult to reduce that uncertainty to
what is possible for $\mu^{\mathrm{(ex)}}$. Comparisons based upon preliminary
trials support the numerical accuracy of the 
evaluations of $h^{\mathrm{(ex)}}$, and hence $c_p^{\mathrm{(ex)}}$. 

The free energy $\mu^{\mathrm{(ex)}}$ increases
with temperature as do the contributions from revised chemistry and
long-range interactions. $\mu^{\mathrm{(ex)}} <
0$ (\textbf{SI}), in contrast with what is observed for CH$_4$.
The revised chemistry contribution is also negative (in
contrast to CH$_4$), indicating that the flooding of the inner shell
with solvent is accompanied by a lowering of the free energy of the
solute. The long-range contribution is negative and 
increases with temperature, just as is found for CH$_4$. The
solvent exclusion contribution for the peptides decreases with increasing
temperature, whereas it increases with increasing temperature for
CH$_4$. That is, the solvent exclusion contribution for a collection of repulsive
cavities (with $\lambda = 3$~{\AA}) of the shape of the peptide
\emph{does not} conform to the behavior expected for a single repulsive
$\lambda = 3$~{\AA} cavity. 

Examining the heat capacity data, we find that for the peptides used in
this study, the backbone solvent contribution makes the largest
contribution, whereas the reorganization contribution is smaller and
similar in magnitude to the contribution from side-chain solvent
interactions. This is in contrast to the behavior of CH$_4$, where
reorganization dominates solute-solvent interactions in $c_p^{(\rm ex)}$.

\subsection*{Hydration of the helix-pair}
\begin{figure*}
\includegraphics[width=0.975\textwidth]{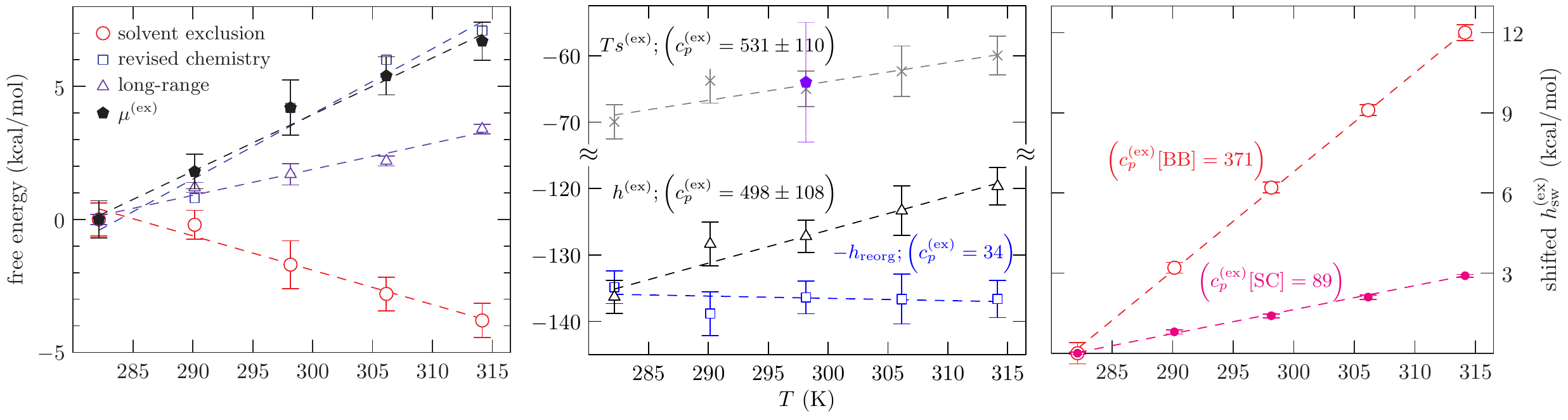} \\
\caption{As in Fig.~\ref{fg:helixcoil}, results for the hydration of
the helix-pair with anti-parallel macro-dipoles.}\label{fg:anti}
\end{figure*}

Fig.~\ref{fg:anti} collects the results on the hydration of the
helix-pair in which the helices are aligned with the macro-dipoles
anti-parallel, as might be found in a helix-turn-helix motif. As before,
the thermodynamic components emphasize the internal consistency of the
data. 

The trends in the data are similar to what one finds for the helix or
coil conformers (Fig.~\ref{fg:helixcoil}). The decrease with
increasing temperature of the solvent exclusion contribution is even more clearly displayed for the
larger helix-pair complex. We can consider the heat capacity change in
the disassembly of the helix-pair to a pair of isolated helices, a
simple model of disassembly of protein complexes. The heat capacity
change, $\Delta c_p^{\mathrm{(ex)}}$, in this process is $32\pm
128$~cal/mol-K --- we use the $c_p^{\mathrm{(ex)}}[h^{\mathrm{(ex)}}]$
values throughout. The uncertainty is necessarily high for reasons noted
above, but focusing on the mean value, the \emph{trends} suggest a
positive contribution, just as was found in early studies on protein
unfolding \cite{privalov:1974kv}.  For the
helix-pair$\rightarrow$helix+helix reaction, $-7$~cal/mol-K is
contributed from backbone solvent contributions, $7$~cal/mol-K from
side-chain solvent contributions, and the remainder from water
reorganization contributions. Note that 
the small net $\Delta c_p^{\mathrm{(ex)}}$ comes from large compensating
physical contributions, and  the agreement of the net $\Delta
c_p^{\mathrm{(ex)}}$ with the solvent reorganization contribution is
fortuitous. Importantly, one cannot make general claims that the
\emph{sign} of $\Delta c_p^{\mathrm{(ex)}}$ is determined by the
reorganization contribution (and hence with models relating to water
structuring around hydrophobic groups). 

It is well appreciated by now that atttactive solute forces are exhibited 
differently in the context of hydrophobic interactions in contrast to 
hydrophobic hydration \cite{pratt1980effects,asthagiri:jcp2008,pratt2016statistical,doi:10.1021/acs.jpcb.8b01711}.

The hydration of the peptide models clearly shows that the observed
$c_p^{\mathrm{(ex)}} > 0 $ and $s^{\mathrm{(ex)}} < 0$ arise from the
attractive protein-solvent contributions to hydration and not from
primitive hydrophobic contributions. What explains the weakening of the
effective protein-solvent attraction with increasing temperature? We
turn to this question next. 

\subsection*{Expansion of the solvent matrix as a basis for understanding the 
temperature signatures} In the temperature range considered here, water expands 
upon heating. Consequently, the inner shell population decreases.
The mean binding energy of the solute is well correlated 
with the number of water molecules in the inner-shell (Fig.~\ref{fg:expansion}). 
This effect is small for a small solute, but is
amplified at the scale of the peptide.  The mean binding energy of the solute 
with the solvent in the inner shell is weaker (or less favorable) at the 
higher temperature.   The
distribution of the binding energy for a given coordination
(\textbf{SI}) supports this suggestion.

\section*{Discussion}
In the hydration of CH$_4$, we find that $s^{\mathrm{(ex)}} < 0$ and
$c_p^{\mathrm{(ex)}} > 0$.  The negative entropy of hydration is often 
interpreted in terms of specific solvent iceberg structures. The positive
$c_p^{\mathrm{(ex)}}$ is then  interpreted as arising from the heat required 
to ``melt'' the ``iceberg.'' 

\begin{figure*}
\centering
\includegraphics[width=0.975\textwidth]{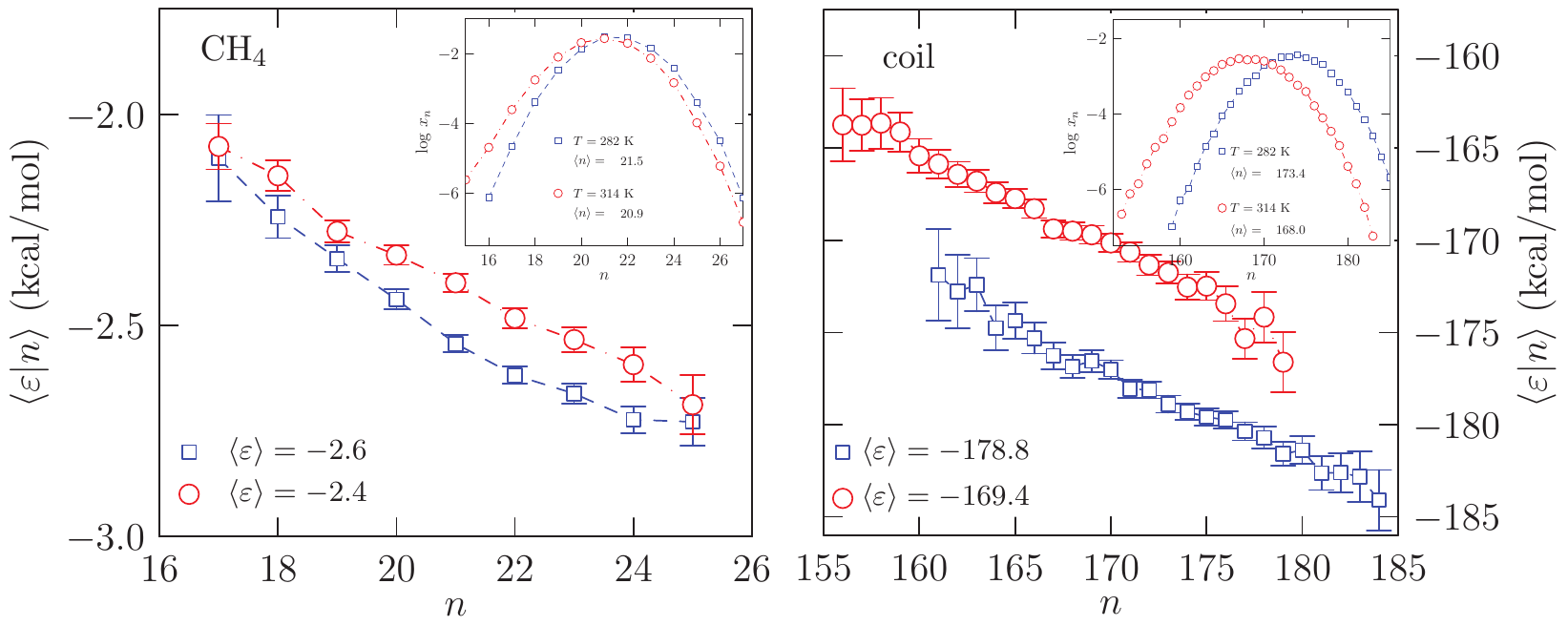} \\
\caption{The conditional mean binding energy of the solute with the solvent in the
inner-shell versus the occupation number $n$ of the inner-shell. Inset:
the probability distribution of coordination states
$\{x_n\}$ versus $n$. Note that the $x_0$ member of this distribution
gives the chemistry contribution [\eqref{eq:qc}]. \underline{Left
panel}: Data for CH$_4$. \underline{Right panel}: Data for the coil
conformer.}\label{fg:expansion}
\end{figure*}

The simple  explanation for these signatures suggested by the present results follows from 
the gradual decrease in solvent population around the solute as the temperature is 
increased. Not only does the population decrease, but for a given population of 
solvent around the solute,  the attractive binding energy of the solvent with the 
solute becomes less favorable (Fig.~\ref{fg:expansion}) as does the interaction between solvent molecules (\textbf{SI}). 
These changes contribute, respectively, to the solute-solvent interaction part of the enthalpy and the reorganization term. This then leads
to $c_p^{\mathrm{(ex)}} > 0$.  Then the  increasing $\mu^{\mathrm{(ex)}}$
with increasing temperature implies $s^{\mathrm{(ex)}}
< 0$.

For the protein, in contrast to what is observed for CH$_4$, the
$s^{\mathrm{(ex)}} < 0$ and $c_p^{\mathrm{(ex)}} > 0$ signatures do not arise from
the temperature dependence of the hydrophobic contribution. 
Thus, the present work shows that attractive protein solvent interactions play an important, even
dominant, role in protein solution thermodynamics. Our work suggests the need
to characterize backbone-solvent interactions, which contribute to the
heat capacity of hydration.  This extends to an explanation for cold denaturation:
with decreasing temperature, the hydration of the unfolded state is
preferred over the folded state because of the favorable hydration of
the peptide backbone. Exploring this hypothesis further is left for
future studies.

\subsection{Methods: Molecular models}

The simulation approach closely follows previous work
\cite{tomar:jpcb16}. The relevant simulation details are summarized in
the Supporting Information.  The deca-alanine peptide was  modeled with an
acetylated (ACE) N-terminus and n-methyl-amide (NME) capped C-terminus.
The extended $\beta$-conformation ($\phi, \psi = -154\pm 12, 149 \pm 9$)
was aligned with the end-to-end vector along the diagonal of
the simulation cell. The single helix
and the helix-pair are aligned with the long axis along the $x$-axis of
the cell. These structures were taken from our earlier work
\cite{tomar:jpcb16}. In the hydration calculation the molecules have a
fixed conformation. The CH$_4$ group is constructed from the CH$_3$
side-chain of the alanine residue. Version C31 of the CHARMM
\cite{charmm} forcefield with correction(cmap) terms for dihedral angles
\cite{cmap2}  was used for the protein models and for CH$_4$. Water
molecules were described by the TIP3P \cite{tip32,tip3mod} designed 
for the CHARMM forcefield.

\begin{acknowledgements}
This research used resources of the National Energy Research Scientific Computing Center, which is supported by the Office of Science of the U.S. Department of Energy under Contract No. DE- AC02-05CH11231. DA thanks Walter G. Chapman for helpful discussions and encouragement. 
\end{acknowledgements}

\end{document}

% \maketitle

\setcounter{section}{0}
\makeatletter
\renewcommand{\thesection}{S.\@arabic\c@section}

\setcounter{figure}{0}
\makeatletter
\renewcommand{\thefigure}{S\@arabic\c@figure}

\setcounter{table}{0}
\makeatletter
\renewcommand{\thetable}{S.\@Roman\c@table}

\setcounter{equation}{0}
\makeatletter
\renewcommand{\theequation}{S.\@arabic\c@equation}

\section{Simulation Details}

The calculation of the hydration free energy using the regularization procedure closely follows earlier studies \cite{tomar:jpcb16,asthagiri:gly15}. We refer the reader to the original articles for complete details. We briefly note the following. 

The chemistry and packing contributions are calculated using thermodynamic integration. To build the field to its eventual range of $\lambda_G = 5$~{\AA}, we progressively apply the field, and for every unit {\AA} increment in the range, we compute the work done in applying the field using a five-point Gauss-Legendre quadrature \cite{Hummer:jcp96}. At each Gauss-point, the system was simulated for $t_{total}$ (time units) and the
last $t_{prod}$ (time units) used for accumulating data. Error analysis and error propagation was performed as before \cite{Weber:jctc12}: the standard error of the mean force was obtained using the Friedberg-Cameron algorithm \cite{friedberg:1970,allen:error} and in adding multiple quantities, the errors were propagated using standard variance-addition rules. 

The starting configuration for each $\lambda$ point is obtained from the ending configuration of the previous point in the chain of states. For the packing contribution, a total of 25 Gauss points span $\lambda \in [0,5]$.  For the chemistry contribution, since solvent never enters $\lambda < 2.9$~{\AA}, conservatively we simulate $\lambda \in [2.5,5]$ for a total of 13 Gauss points. 

For CH$_4$, the long-range contribution was obtained using the inverse form of the potential distribution theorem in the presence of the external field with range $\lambda_G = 5$~{\AA}. In this case, from the end-point of the chemistry calculation, we create the system with $\lambda_G = 5$~{\AA}. The system was simulated for 
a total $t_{lr,total}$ (time units) and solute-solvent interaction accumulated over $t_{lr,prod}$ (time units).   

For the peptide models, the long-range contribution was obtained using the forward form of the potential distribution theorem in the presence of the external field with range $\lambda_G = 5$~{\AA}. In this case, from the end-point of the packing calculation, we create the system with $\lambda_G = 5$~{\AA}. The system was simulated for 
a total $t_{lr,total}$ (time units) and neat solvent frames archived over $t_{lr,prod}$ (time units).  We then do test particle insertions in the neat solvent frame to obtain the solute-solvent binding energy distribution.  

Specifics for each system is given below. 
\subsection{CH$_4$}
The water box comprises 992 TIP3P\cite{tip32,tip3mod} water molecules. For the packing and chemistry calculations, at each $\lambda$, $t_{total} = 4$~ns and $t_{prod} = 3$~ns. Frames were 
archived every 100~fs for further analysis. For the long-range calculation, the system with $\lambda_G = 5$~{\AA} was simulated for 5~ns and frames archived every 400~fs for analysis. 

For the excess enthalpy and coordination number analysis, the well-equilibrated solute-solvent system (with no external field) was further simulated for 5~ns and frames archived every 200~fs for analysis.  

\subsection{Helix and coil conformers}
As before \cite{tomar:jpcb16} we used a box of 3500 TIP3P water molecules. For the packing and chemistry calculations, at each $\lambda$, $t_{total} = 1$~ns and $t_{prod} = 0.75$~ns. (We checked to ensure that neglecting more data does not perceptibly change the results.) Frames were archived every 50~fs for further analysis. (For the 298.15~K case alone, 
the simulations were repeated 4 times and the results averaged. We emphasize that the results from the four independent repeats were consistent within the statistical uncertainty of each run.) For the long-range calculation, the system with $\lambda_G = 5$ was simulated for $t_{lr,total} = 4$~ns. Data was accumulated over the last $t_{lr,prod} = 2$~ns, with frames saved every 500~fs. For temperatures except 298.15~K, this procedure was repeated twice. The resulting data was pooled together and analyzed as before \cite{tomar:jpcb16,asthagiri:gly15}. For the 298.15~K case, for the outer calculation we simulated for $t_{tr,total} = 11$~ns and accumulated data for $t_{lr,prod} = 10$~ns, with frames saved every 500~fs. This procedure was repeated twice and the results pooled and analyzed. We emphasize that the Gaussian description of shorter segments of the long 10~ns production in the 298.15~K case was in excellent agreement with the Gaussian description using all the data. This confirms the robustness of the Gaussian and 
shows that our choices for the other temperatures are still very conservative. 

For the excess enthalpy and coordination number analysis, the well-equilibrated solute-solvent system (with no external field) was simulated for 4~ns and frames archived every 200~fs for analysis. 

\subsection{Helix-pairs}

For the helix-pairs, the chemistry and packing calculation was as for the isolated helix. For the long-range calculation, for the 298.15~K case, the system was equilibrated for 1~ns. Five independent 2~ns runs were performed and the data analyzed. For the rest of the temperatures, $t_{lr,total} = 11$~ns of which $t_{lr,prod} = 10$~ns was used for accumulating data. Frames were saved at the same rate as for the isolated helix. 

For the excess enthalpy calculation, the well-equilibrated solute-solvent system (with no external field) was simulated for a total of 3~ns and the last 2.75~ns used for analysis. Frames were archived every 500~fs for a total of 5500 frames. Because of a factor of 3 less data (compared to the isolated protein case), the statistical uncertainties were correspondingly higher. 

\section{Entropy from the temperature derivative of $\mu^{(\mathrm ex)}$}

To check the consistency of $Ts^{\mathrm{(ex)}}$ calculated at $T = 298.15$~K using Euler's relation (Eq.~3, main text) with $Ts^{\mathrm{(ex)}}  = -T(\partial \mu^{\mathrm{(ex)}} / \partial T)_{N,p}$, 
we can either take the derivative of $\mu^{\mathrm{(ex)}}$ using a central difference formula or by differentiating a model fit to the calculated $\mu^{\mathrm{(ex)}}$ values. Statistically, the latter approach is to be preferred given the logarithmic dependence of $\mu^{\mathrm{(ex)}}$ on $T$. To this end, consistent with the analysis of $c_p^{\mathrm{(ex)}}$ (main text), we assume constant heat capacity over the temperature range of interest and set $\mu^{(\mathrm ex)}_{model} = a + b \cdot T + c \cdot T \log T$ with $c < 0$.  To estimate the model parameters $a, b, c$, we minimize
\begin{eqnarray}
\chi^2 = \sum\limits_{i} \left( \frac{ \mu^{(\mathrm ex)}_{\rm calc}(T_i) - \mu^{(\mathrm ex)}_{\rm model}(T_i) }{\bar{\sigma}(T_i)} \right)^2 \, ,
\label{eq:s1}
\end{eqnarray}
using the interior point method within Mathematica\cite{Mathematica}. In Eq.~\ref{eq:s1}, $\mu^{(\mathrm ex)}_{\rm calc}(T_i)$ is the calculated value of $\mu^{(\mathrm ex)}$ at the temperature $T_i$ and  $\bar{\sigma}(T_i)$ is the associated standard error of the mean value.  To estimate the uncertainty of $Ts^{\mathrm{(ex)}}$ from the derivative path, we 
repeated the above minimization procedure for synthetic data sets obtained by drawing random variates using $\mu^{(\mathrm ex)}_{\rm calc}(T_i)$ and ${\bar{\sigma}(T_i)}$ as the mean and standard deviation. We repeated this simulation procedure over 100 times and found the numerical estimates to be well-converged. (In practice, as few as 10 data sets suffice to obtain a robust estimate of the mean value.) The standard deviation of the distribution of  $Ts^{\mathrm{(ex)}}$ values calculated over the synthetic data set is taken as the estimate of the uncertainty of $Ts^{\mathrm{(ex)}}_{model}$ calculated from the derivative of $\mu^{\mathrm{(ex)}}_{model}$. 

Table~\ref{tb:derivatives} below collects the results from the above analysis. Please note that $c^{\mathrm{(ex)}}_{p,model}$ is obtained from a second derivative of $\mu^{(\mathrm ex)}_{model}$ with respect to temperature, and not surprisingly, the variation is rather large. For comparison, we reproduce $Ts^{\mathrm{(ex)}}$ from Euler's relation (Eq.~3 main text) and $c_p^{\mathrm{(ex)}}$ from the temperature derivative of $h^{\mathrm{(ex)}}$. 
\ctable[
	mincapwidth=\textwidth,
	caption={$Ts^{\mathrm{(ex)}}_{model}$ (in kcal/mol) at $T=298.15$~K and $c^{\mathrm{(ex)}}_{p,model}$ (in cal/mol-K). For $c^{\mathrm{(ex)}}_{p,model}$, the numbers in parenthesis provide the range containing 90\% of the estimated model values.}, 
	label=tb:derivatives,
	pos=h,
	captionskip=-1.5ex
	]
{c r r r r}

{
\FL
Species & $Ts^{\mathrm{(ex)}}_{model}$ & $Ts^{\mathrm{(ex)}}_{Euler}$ & $c^{\mathrm{(ex)}}_{p,model}$  & $c^{\mathrm{(ex)}}_{p} [h^{\mathrm{(ex)}}]$ \ML
CH$_4$ & $  -2.7 \pm  0.5 $  & $ -3.6 \pm  0.4 $  & $  16\, (0,60) $  & $ 31 \pm 4 $   \NN[-1ex]
Helix      & $  -39 \pm 5 $  & $ -40 \pm  1 $  & $  1198\,  (0,2480) $  & $ 265 \pm  48 $   \NN[-1ex]
Coil & $ -54 \pm  4 $  & $ -54 \pm  2 $  & $  1278\, (0, 2120) $  & $ 340 \pm  82 $  \NN[-1ex]
Helix pair (antiparallel)  & $  -64 \pm  9 $  & $ -65 \pm 3 $  & $  1585\, (0, 3385) $  & $ 498 \pm 108 $   \NN[-1ex]
Helix pair (parallel) & $ -71 \pm 11 $  & $ -71 \pm 3 $  & $  1866\, (0, 4270) $  & $ 541 \pm 69 $   \LL
}

% \newpage

\section{Reorganization energy by shells at 298.15~K}

The calculation of the reorganization contribution follows the procedure exhaustively documented before \cite{tomar:jpcb16,asthagiri:gly15}. Please note that  we define an inner-shell around the peptide as the union of shells of radius $\lambda$ centered  on the peptide heavy atoms.  $\lambda \leq 5.5$~{\AA}, $5.5 < \lambda \leq 8.5$~{\AA}, and $8.5 < \lambda \leq 11.5$~{\AA}  defined the first, second, and third shells, respectively. For the reorganization calculation, the definition of the inner shell was slightly increased by 0.5~{\AA}, but this change has no bearing on the final thermodynamic quantity $h^{(\rm ex)}$. 
\begin{figure}[h!]
\includegraphics[width=3.25in]{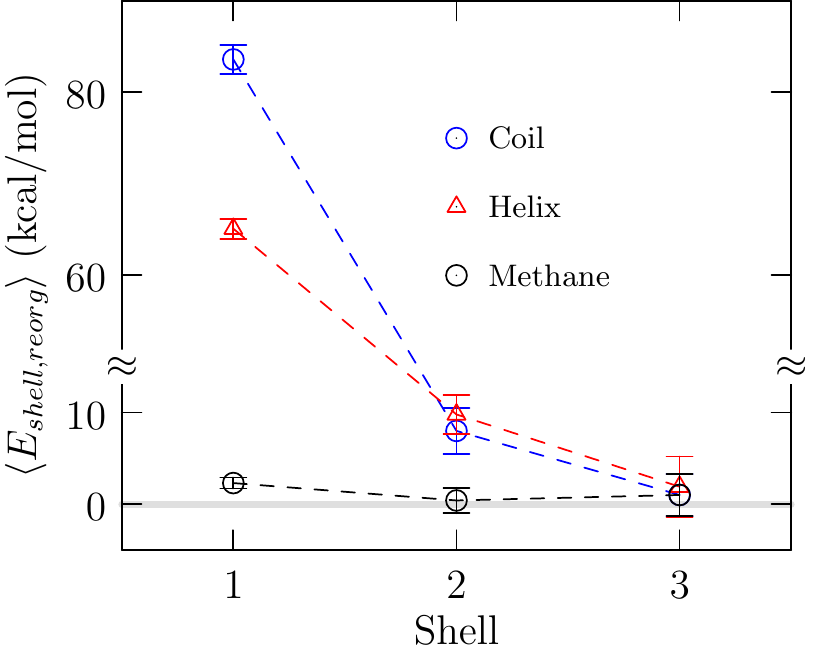}
\caption{Convergence of reorganization energy as a function of the shells of water around the solute. For methane the reorganization contribution converges within the first shell, whereas we require the first two shells for the helix and coil conformers. The same holds true for the helix-pair.}\label{fg:reorg}
\end{figure}

\newpage

\section{Hydration of the helix-pair with helix macro-dipoles oriented in the parallel configuration.}

\begin{figure}[h!]
\includegraphics[width=0.95\textwidth]{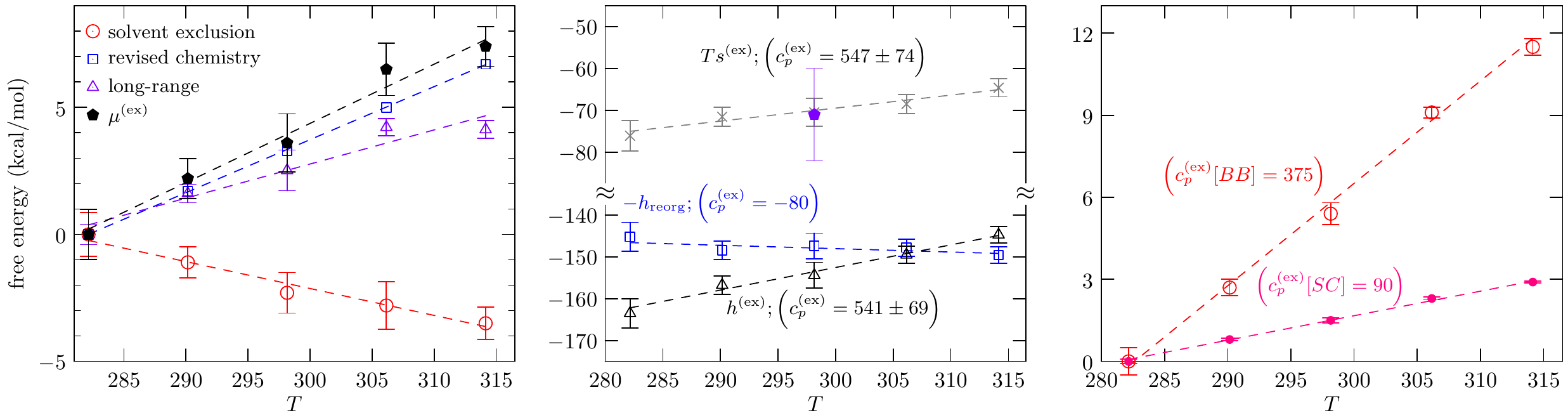}
\caption{As in Figure 4 (main text) for a helix-pair, but with helices in a parallel configuration in the helix pair.}
\end{figure}

\section{Effect of expansion of the matrix}
\begin{figure}[h!]
\includegraphics[scale=0.95]{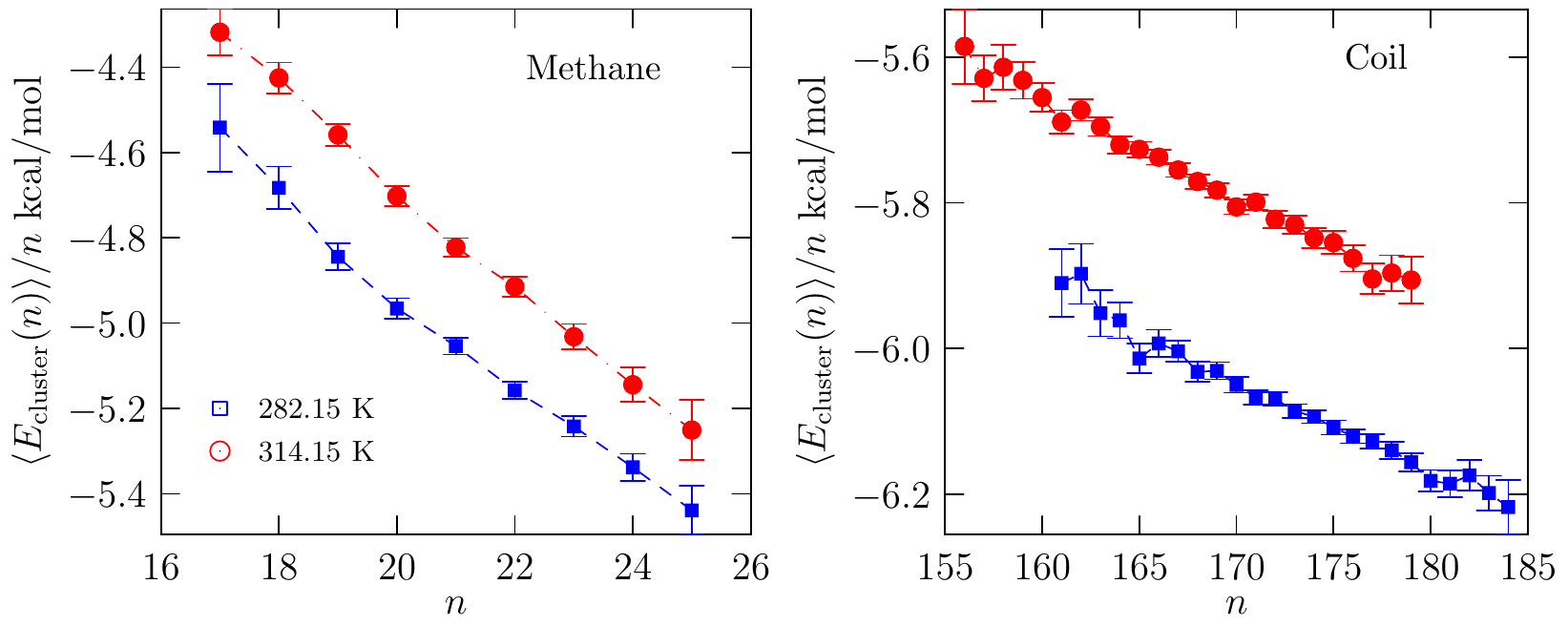}
\caption{Variation of the energy of the water cluster in the inner shell. Please note that the cluster energy with the rest of the bulk solvent or the solute
is not taken into consideration.}\label{fg:clus}
\end{figure}

\begin{figure}[h!]
\includegraphics[scale=0.95]{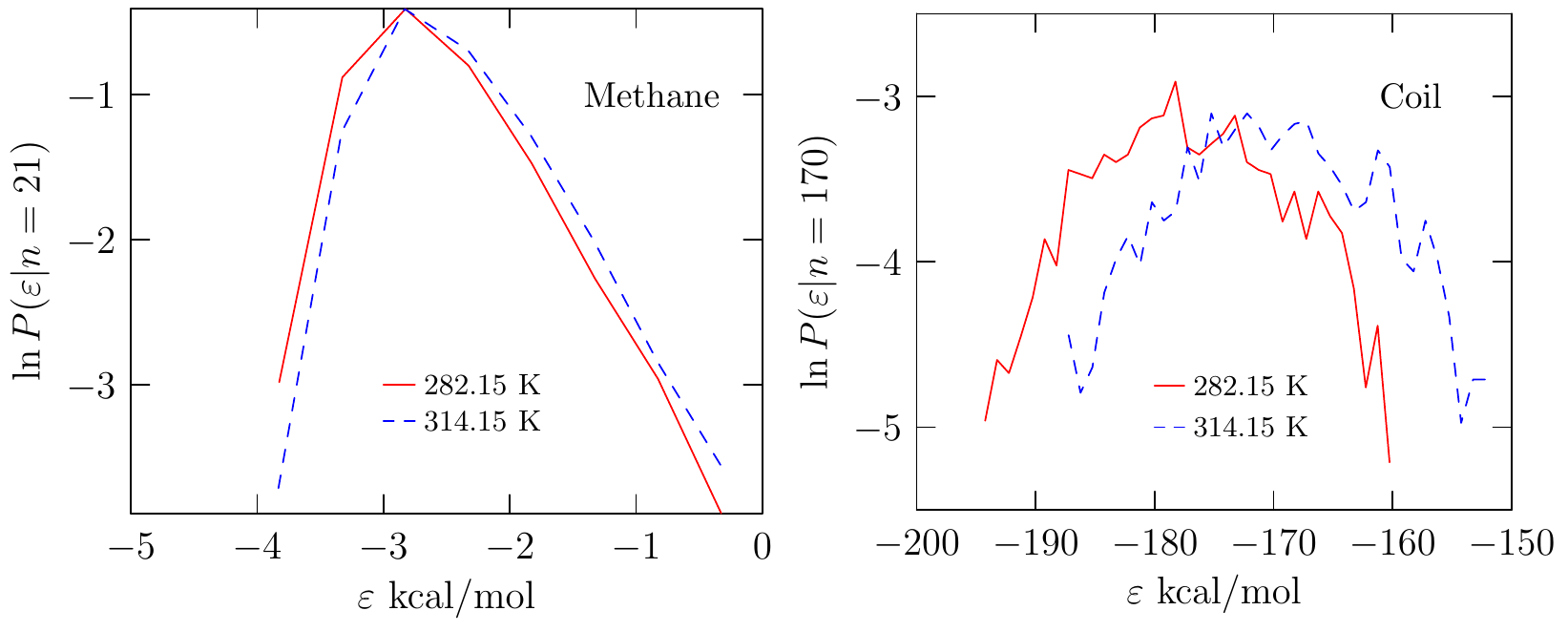}
\caption{Figure showing the distribution of solute interaction with solvent in the inner shell for a defined coordination $n$ of the inner shell. Observe that both the low energy tail and the high energy tails are affected by increase in temperature. This suggests that the solvent samples  both the repulsive wall and the weaker attractive tail as temperature is increased.}\label{fg:be}
\end{figure}

\newpage

\section{Hydration of CH$_4$}
\ctable[
	mincapwidth=\textwidth,
	caption={Quasicomponents}, 
	label=tb:ch4qc,
	pos=h,
	captionskip=-1.5ex
	]
{c r r r r}

{
\FL
Temperature & Solvent exclusion & Revised Chemistry & Long-range & $\mu^{(\rm ex)}$ \ML
282.15 & $  2.47 \pm  0.01 $  & $  2.00 \pm  0.05 $  & $ -2.21 \pm  0.003 $  & $  2.26 \pm  0.06 $ \NN[-1ex]
290.15 & $  2.48 \pm  0.01 $  & $  1.98 \pm  0.04 $  & $ -2.18 \pm  0.003 $  & $  2.28 \pm  0.04 $ \NN[-1ex]
298.15 & $  2.49 \pm  0.01 $  & $  2.01 \pm  0.05 $  & $ -2.15 \pm  0.003 $  & $  2.35 \pm  0.05 $ \NN[-1ex]
306.15 & $  2.52 \pm  0.01 $  & $  2.06 \pm  0.05 $  & $ -2.13 \pm  0.003 $  & $  2.45 \pm  0.05 $ \NN[-1ex]
314.15 & $  2.53 \pm  0.01 $  & $  2.08 \pm  0.04 $  & $ -2.09 \pm  0.003 $  & $  2.52 \pm  0.04 $ \LL
}

\ctable[
	mincapwidth=\textwidth,
	caption={Thermodynamic components}, 
	label=tb:ch4qc,
	pos=h,
	captionskip=-1.5ex
	]
{c r r r r}

{
\FL
Temperature & $h^{(\rm ex)}_{sw}$ & $h_{reorg}$ & $h^{(\rm ex)}$ & $Ts^{(\rm ex)}$  \ML
282.15 & $ -3.39 \pm  0.01 $  & $  1.80 \pm  0.30 $  & $ -1.60 \pm  0.30 $  & $ -3.80 \pm  0.30 $ \NN[-1ex]
290.15 & $ -3.34 \pm  0.01 $  & $  2.00 \pm  0.40 $  & $ -1.40 \pm  0.40 $  & $ -3.60 \pm  0.40 $ \NN[-1ex]
298.15 & $ -3.30 \pm  0.01 $  & $  2.10 \pm  0.40 $  & $ -1.20 \pm  0.40 $  & $ -3.60 \pm  0.40 $ \NN[-1ex]
306.15 & $ -3.25 \pm  0.01 $  & $  2.50 \pm  0.40 $  & $ -0.80 \pm  0.40 $  & $ -3.20 \pm  0.40 $ \NN[-1ex]
314.15 & $ -3.21 \pm  0.01 $  & $  2.60 \pm  0.30 $  & $ -0.60 \pm  0.30 $  & $ -3.10 \pm  0.30 $ \LL
}

\newpage

\section{Hydration of deca-alanine helix}

\ctable[
	mincapwidth=\textwidth,
	caption={Quasicomponents}, 
	label=tb:ch4qc,
	pos=h,
	captionskip=-1.5ex
	]
{c r r r r}

{
\FL
Temperature & Solvent exclusion & Revised Chemistry & Long-range & $\mu^{(\rm ex)}$ \ML
282.15 & $  44.7 \pm   0.5 $  & $ -53.2 \pm   0.7 $  & $ -32.8 \pm   0.2 $  & $ -41.4 \pm   0.6 $ \NN[-1ex]
290.15 & $  45.3 \pm   0.3 $  & $ -52.9 \pm   0.5 $  & $ -32.1 \pm   0.2 $  & $ -39.7 \pm   0.4 $ \NN[-1ex]
298.15 & $  44.4 \pm   0.4 $  & $ -51.6 \pm   0.6 $  & $ -31.6 \pm   0.1 $  & $ -38.9 \pm   0.5 $ \NN[-1ex]
306.15 & $  44.2 \pm   0.3 $  & $ -51.1 \pm   0.4 $  & $ -30.9 \pm   0.2 $  & $ -37.8 \pm   0.4 $ \NN[-1ex]
314.15 & $  43.9 \pm   0.3 $  & $ -50.2 \pm   0.4 $  & $ -30.8 \pm   0.2 $  & $ -37.1 \pm   0.4 $ \LL
}

\ctable[
	mincapwidth=\textwidth,
	caption={Thermodynamic components}, 
	label=tb:ch4qc,
	pos=h,
	captionskip=-1.5ex
	]
{c r r r r r}

{
\FL
Temperature & $h^{(\rm ex)}_{bb}$ & $h^{(\rm ex)}_{sc}$  & $h_{reorg}$ & $h^{(\rm ex)}$ & $Ts^{(\rm ex)}$  \ML
282.15 & $ -129.0 \pm   0.1 $  & $ -26.10 \pm  0.02 $  & $  71.3 \pm   1.8 $  & $ -83.8 \pm   1.8 $  & $ -42.5 \pm   1.9 $ \NN[-1ex]
290.15 & $ -127.5 \pm   0.1 $  & $ -25.80 \pm  0.02 $  & $  72.7 \pm   1.8 $  & $ -80.6 \pm   1.9 $  & $ -40.9 \pm   1.9 $ \NN[-1ex]
298.15 & $ -125.9 \pm   0.1 $  & $ -25.30 \pm  0.02 $  & $  72.8 \pm   1.3 $  & $ -78.4 \pm   1.3 $  & $ -39.6 \pm   1.4 $ \NN[-1ex]
306.15 & $ -124.6 \pm   0.1 $  & $ -25.01 \pm  0.03 $  & $  72.7 \pm   2.2 $  & $ -76.9 \pm   2.2 $  & $ -39.1 \pm   2.2 $ \NN[-1ex]
314.15 & $ -123.2 \pm   0.1 $  & $ -24.60 \pm  0.02 $  & $  72.6 \pm   2.0 $  & $ -75.2 \pm   2.0 $  & $ -38.1 \pm   2.0 $ \LL
}

\newpage

\section{Hydration of deca-alanine coil $C_0$}

\ctable[
	mincapwidth=\textwidth,
	caption={Quasicomponents}, 
	label=tb:ch4qc,
	pos=h,
	captionskip=-1.5ex
	]
{c r r r r}

{
\FL
Temperature & Solvent exclusion & Revised Chemistry & Long-range & $\mu^{(\rm ex)}$ \ML
282.15 & $  58.4 \pm   0.1 $  & $ -79.0 \pm   0.3 $  & $ -28.8 \pm   0.1 $  & $ -49.4 \pm   0.3 $ \NN[-1ex]
290.15 & $  58.3 \pm   0.2 $  & $ -77.8 \pm   0.4 $  & $ -28.1 \pm   0.1 $  & $ -47.7 \pm   0.4 $ \NN[-1ex]
298.15 & $  58.4 \pm   0.2 $  & $ -77.0 \pm   0.4 $  & $ -27.5 \pm   0.0 $  & $ -46.1 \pm   0.3 $ \NN[-1ex]
306.15 & $  58.2 \pm   0.2 $  & $ -75.7 \pm   0.4 $  & $ -27.0 \pm   0.1 $  & $ -44.5 \pm   0.3 $ \NN[-1ex]
314.15 & $  57.3 \pm   0.2 $  & $ -74.8 \pm   0.3 $  & $ -26.2 \pm   0.1 $  & $ -43.7 \pm   0.3 $ \LL
}

\ctable[
	mincapwidth=\textwidth,
	caption={Thermodynamic components}, 
	label=tb:ch4qc,
	pos=h,
	captionskip=-1.5ex
	]
{c r r r r r}

{
\FL
Temperature & $h^{(\rm ex)}_{bb}$ & $h^{(\rm ex)}_{sc}$  & $h_{reorg}$ & $h^{(\rm ex)}$ & $Ts^{(\rm ex)}$  \ML
282.15 & $ -163.3 \pm   0.1 $  & $ -28.00 \pm  0.02 $  & $  88.5 \pm   1.7 $  & $ -102.8 \pm   1.7 $  & $ -53.5 \pm   1.7 $ \NN[-1ex]
290.15 & $ -161.6 \pm   0.1 $  & $ -27.60 \pm  0.02 $  & $  90.2 \pm   1.7 $  & $ -99.0 \pm   1.8 $  & $ -51.3 \pm   1.8 $ \NN[-1ex]
298.15 & $ -159.5 \pm   0.2 $  & $ -27.20 \pm  0.02 $  & $  87.1 \pm   2.0 $  & $ -99.6 \pm   2.0 $  & $ -53.5 \pm   2.1 $ \NN[-1ex]
306.15 & $ -157.6 \pm   0.1 $  & $ -26.90 \pm  0.03 $  & $  90.0 \pm   2.1 $  & $ -94.5 \pm   2.1 $  & $ -50.0 \pm   2.1 $ \NN[-1ex]
314.15 & $ -155.5 \pm   0.2 $  & $ -26.40 \pm  0.02 $  & $  90.3 \pm   1.5 $  & $ -91.6 \pm   1.5 $  & $ -47.9 \pm   1.5 $ \LL
}

\newpage
\section{Hydration of helix pair with helix macro-dipoles in the antiparallel configuration}

\ctable[
	mincapwidth=\textwidth,
	caption={Quasicomponents}, 
	label=tb:ch4qc,
	pos=h,
	captionskip=-1.5ex
	]
{c r r r r}

{
\FL
Temperature & Solvent exclusion & Revised Chemistry & Long-range & $\mu^{(\rm ex)}$ \ML
282.15 & $  84.1 \pm   0.6 $  & $ -107.6 \pm   0.9 $  & $ -42.9 \pm   0.2 $  & $ -66.4 \pm   0.7 $ \NN[-1ex]
290.15 & $  83.9 \pm   0.5 $  & $ -106.8 \pm   0.8 $  & $ -41.7 \pm   0.2 $  & $ -64.6 \pm   0.6 $ \NN[-1ex]
298.15 & $  82.4 \pm   0.9 $  & $ -103.4 \pm   1.3 $  & $ -41.2 \pm   0.4 $  & $ -62.2 \pm   1.0 $ \NN[-1ex]
306.15 & $  81.3 \pm   0.6 $  & $ -101.6 \pm   0.9 $  & $ -40.7 \pm   0.2 $  & $ -61.0 \pm   0.7 $ \NN[-1ex]
314.15 & $  80.3 \pm   0.6 $  & $ -100.5 \pm   0.9 $  & $ -39.5 \pm   0.2 $  & $ -59.7 \pm   0.7 $ \LL
}
\ctable[
	mincapwidth=\textwidth,
	caption={Thermodynamic components }, 
	label=tb:ch4qc,
	pos=h,
	captionskip=-1.5ex
	]
{c r r r r r}

{
\FL
Temperature & $h^{(\rm ex)}_{bb}$ & $h^{(\rm ex)}_{sc}$  & $h_{reorg}$ & $h^{(\rm ex)}$ & $Ts^{(\rm ex)}$  \ML
282.15 & $ -226.0 \pm   0.4 $  & $ -45.20 \pm  0.04 $  & $ 134.9 \pm   2.5 $  & $ -136.3 \pm   2.5 $  & $ -69.9 \pm   2.6 $ \NN[-1ex]
290.15 & $ -222.8 \pm   0.2 $  & $ -44.40 \pm  0.08 $  & $ 138.8 \pm   3.3 $  & $ -128.4 \pm   3.3 $  & $ -63.8 \pm   3.4 $ \NN[-1ex]
298.15 & $ -219.8 \pm   0.2 $  & $ -43.80 \pm  0.07 $  & $ 136.4 \pm   2.4 $  & $ -127.2 \pm   2.4 $  & $ -65.0 \pm   2.7 $ \NN[-1ex]
306.15 & $ -216.9 \pm   0.2 $  & $ -43.10 \pm  0.08 $  & $ 136.7 \pm   3.7 $  & $ -123.3 \pm   3.7 $  & $ -62.4 \pm   3.8 $ \NN[-1ex]
314.15 & $ -214.0 \pm   0.3 $  & $ -42.30 \pm  0.05 $  & $ 136.6 \pm   2.8 $  & $ -119.7 \pm   2.8 $  & $ -60.0 \pm   2.9 $ \LL
}

\newpage
\section{Hydration of helix pair with helix macro-dipoles in the parallel configuration}

\ctable[
	mincapwidth=\textwidth,
	caption={Quasicomponents}, 
	label=tb:ch4qc,
	pos=h,
	captionskip=-1.5ex
	]
{c r r r r}

{
\FL
Temperature & Solvent exclusion & Revised Chemistry & Long-range & $\mu^{(\rm ex)}$ \ML
282.15 & $  84.3 \pm   0.9 $  & $ -109.1 \pm   1.2 $  & $ -62.7 \pm   0.4 $  & $ -87.5 \pm   1.0 $ \NN[-1ex]
290.15 & $  83.2 \pm   0.6 $  & $ -107.4 \pm   0.9 $  & $ -61.1 \pm   0.4 $  & $ -85.3 \pm   0.8 $ \NN[-1ex]
298.15 & $  82.0 \pm   0.8 $  & $ -105.8 \pm   1.1 $  & $ -60.1 \pm   0.8 $  & $ -83.9 \pm   1.1 $ \NN[-1ex]
306.15 & $  81.5 \pm   0.9 $  & $ -104.1 \pm   1.4 $  & $ -58.4 \pm   0.3 $  & $ -81.0 \pm   1.0 $ \NN[-1ex]
314.15 & $  80.8 \pm   0.6 $  & $ -102.4 \pm   0.9 $  & $ -58.5 \pm   0.4 $  & $ -80.1 \pm   0.8 $ \LL
}

\ctable[
	mincapwidth=\textwidth,
	caption={Thermodynamic components}, 
	label=tb:ch4qc,
	pos=h,
	captionskip=-1.5ex
	]
{c r r r r r}

{
\FL
Temperature & $h^{(\rm ex)}_{bb}$ & $h^{(\rm ex)}_{sc}$  & $h_{reorg}$ & $h^{(\rm ex)}$ & $Ts^{(\rm ex)}$  \ML
282.15 & $ -260.0 \pm   0.5 $  & $ -48.70 \pm  0.08 $  & $ 145.2 \pm   3.4 $  & $ -163.5 \pm   3.5 $  & $ -76.0 \pm   3.6 $ \NN[-1ex]
290.15 & $ -257.3 \pm   0.3 $  & $ -47.90 \pm  0.05 $  & $ 148.4 \pm   2.2 $  & $ -156.8 \pm   2.2 $  & $ -71.5 \pm   2.3 $ \NN[-1ex]
298.15 & $ -254.6 \pm   0.4 $  & $ -47.20 \pm  0.09 $  & $ 147.4 \pm   3.1 $  & $ -154.4 \pm   3.1 $  & $ -70.5 \pm   3.3 $ \NN[-1ex]
306.15 & $ -250.9 \pm   0.2 $  & $ -46.40 \pm  0.05 $  & $ 147.8 \pm   2.0 $  & $ -149.5 \pm   2.0 $  & $ -68.5 \pm   2.3 $ \NN[-1ex]
314.15 & $ -248.5 \pm   0.3 $  & $ -45.80 \pm  0.04 $  & $ 149.6 \pm   2.0 $  & $ -144.7 \pm   2.0 $  & $ -64.6 \pm   2.2 $ \LL
}

\newpage

% \bibliography{refs}

\providecommand{\latin}[1]{#1}
\makeatletter
\providecommand{\doi}
  {\begingroup\let\do\@makeother\dospecials
  \catcode`\{=1 \catcode`\}=2 \doi@aux}
\providecommand{\doi@aux}[1]{\endgroup\texttt{#1}}
\makeatother
\providecommand*\mcitethebibliography{\thebibliography}
\csname @ifundefined\endcsname{endmcitethebibliography}
  {\let\endmcitethebibliography\endthebibliography}{}